\documentclass{article} 
\usepackage{gem_workshop_2024,times}

\usepackage[utf8]{inputenc} 
\usepackage[T1]{fontenc}    
\usepackage{hyperref}       
\usepackage{url}            
\usepackage{booktabs}       
\usepackage{amsfonts}       
\usepackage{nicefrac}       
\usepackage{microtype}      
\usepackage{xcolor}         

\usepackage{amsmath,amsfonts,bm}

\def\eqref#1{equation~\ref{#1}}

\def\1{\bm{1}}

\DeclareMathAlphabet{\mathsfit}{\encodingdefault}{\sfdefault}{m}{sl}
\SetMathAlphabet{\mathsfit}{bold}{\encodingdefault}{\sfdefault}{bx}{n}

\def\gL{{\mathcal{L}}}

\def\gR{{\mathcal{R}}}

\newcommand{\E}{\mathbb{E}}

\usepackage{enumitem}
\usepackage{xspace}
\usepackage{xcolor}
\usepackage{wrapfig}
\usepackage{caption}
\usepackage{amssymb}
\usepackage{multirow}
\usepackage{adjustbox}
\usepackage{makecell}
\usepackage{caption}
\usepackage{subcaption}
\usepackage{graphicx}
\usepackage{mathtools}
\usepackage{algorithm}
\usepackage{algpseudocode}
\usepackage{colortbl}
\usepackage{tcolorbox}
\usepackage{babel}
\usepackage{bbding}
\usepackage{pifont}
\usepackage{nicematrix}
\usepackage{enumitem}
\usepackage{booktabs}
\usepackage[flushleft]{threeparttable}

\usepackage{thmtools}
\usepackage{thm-restate}

\definecolor{myblue}{RGB}{68, 114, 196}
\definecolor{myorange}{RGB}{237, 125, 49}
\definecolor{lightgray}{RGB}{229, 232, 232}
\definecolor{mydarkblue}{rgb}{0,0.08,0.45}

\usepackage{hyperref}
\usepackage{url}

\title{Structure-Informed Protein Language Model}

\author{Zuobai Zhang\textsuperscript{1,2}, Jiarui Lu\textsuperscript{1,2}, Vijil Chenthamarakshan\textsuperscript{3},\\ \textbf{Aur\'{e}lie Lozano\textsuperscript{3}, Payel Das\textsuperscript{3}, Jian Tang\textsuperscript{1,4,5}} \\
  Mila - Qu\'ebec AI Institute\textsuperscript{1}, Universit\'e de Montr\'eal\textsuperscript{2}\\
  IBM Research\textsuperscript{3},
  HEC Montr\'eal\textsuperscript{4}, CIFAR AI Chair\textsuperscript{5} \\
  \texttt{\{zuobai.zhang, jiarui.lu\}@mila.quebec}, \\
  \texttt{\{aclozano,ecvijil,daspa\}@us.ibm.com},
  \texttt{jian.tang@hec.ca}
}

\iclrfinalcopy 
\begin{document}

\maketitle

\begin{abstract}
Protein language models are a powerful tool for learning protein representations through pre-training on vast protein sequence datasets. 
However, traditional protein language models lack explicit structural supervision, despite its relevance to protein function. 
To address this issue, we introduce the integration of remote homology detection to distill structural information into protein language models without requiring explicit protein structures as input. 
We evaluate the impact of this structure-informed training on downstream protein function prediction tasks. 
Experimental results reveal consistent improvements in function annotation accuracy for EC number and GO term prediction. Performance on mutant datasets, however, varies based on the relationship between targeted properties and protein structures. This underscores the importance of considering this relationship when applying structure-aware training to protein function prediction tasks. Code and model weights are available at \url{https://github.com/DeepGraphLearning/esm-s}.
\end{abstract}
\section{Introduction}

Proteins play a fundamental role in biological processes, and a deeper understanding of them can pave the way for groundbreaking advancements in medical, pharmaceutical, and genetic research. 
A cutting-edge technology that has emerged to represent proteins is the Protein Language Model (PLM)~\citep{rao2019evaluating}.
Inspired by Natural Language Processing (NLP) methodologies, PLMs have demonstrated remarkable performance in capturing long-range residue correlations - also known as co-evolution - through self-supervised training on vast repositories of protein residue sequences~\citep{rao2021msa}.
Prominent PLMs like ESM~\citep{rives2021esm1b,lin2023evolutionary} have shown the ability to implicitly capture evolutionary and structural information and demonstrate outstanding performance across various tasks related to protein structures and functions.

However, vanilla PLMs face a significant limitation: their absence of explicit supervision based on protein structure information, despite its critical relevance to protein function. 
To address this limitation, recent studies have developed models that combine large-scale pre-training on protein sequences with the integration of structural information as input~\citep{zhang2023enhancing,su2023saprot}. 
While these models have demonstrated impressive performance in function prediction, their reliance on protein structures as input introduces an additional computational burden for structure prediction and limits their application to proteins with indistinct structures, such as mutant data. 
Therefore, a key question remains: how to distill structural knowledge into protein language models without requiring explicit structure as input, and how will this impact downstream function prediction tasks?

In this study, we investigate the use of remote homology detection tasks as a means of incorporating structural information into protein language models~\citep{chen2018comprehensive}. This task aims to identify proteins with similar structures but low sequence similarity, thus complementing the training of protein language models. We train ESM-2 models \citep{lin2023evolutionary} on this task and obtain \emph{structure-informed protein language models}.
To assess the impact of structurally training, we evaluate our models on downstream function prediction tasks taken from \citet{gligorijevic2021structure}, \citet{xu2022peer}, and \citet{dallago2021flip}. We find that incorporating structural information leads to consistent improvement in function annotation tasks like Enzyme Commission (EC) number and Gene Ontology (GO) term prediction. However, improvement on mutant data highly depends on the targeted property's relationship to protein structures.
This highlights the importance of considering the relationship between protein structures and targeted properties when applying structure-informed training to protein function prediction tasks. 
We hope this study encourages further exploration of structural knowledge in protein language models, leading to better protein representation learning.

\textbf{Related work.}
Protein language models view protein sequences as the language of life and employ masked language modeling loss for pre-training transformer-based models~\citep{tape2019, elnaggar2020prottrans, rives2021biological, elnaggar2023ankh}. Structure-based representation learning has also shown promise in incorporating structural information for better representation learning methods~\citep{jing2021equivariant, zhang2022protein, chen2022structure, zhang2023physics}.
To combine the advantages of both approaches, recent studies have designed architectures that take both sequences and structures as input~\citep{zhang2023enhancing, su2023saprot}. However, these methods require protein structures as input, limiting their application to datasets without protein structures.
In this study, we propose the use of remote homology detection to inject structural information into protein language models. While this concept has been utilized in previous works~\citep{bepler2018learning, hamamsy2022tm}, their focus is on structural alignment rather than representation learning. Here, we investigate the impact of incorporating structural information on learning protein representations by evaluating our models on downstream function prediction tasks.

\section{Method}

\begin{figure}[t]
\centering
\begin{subfigure}[b]{0.48\textwidth}
    \centering
    \includegraphics[width=\linewidth]{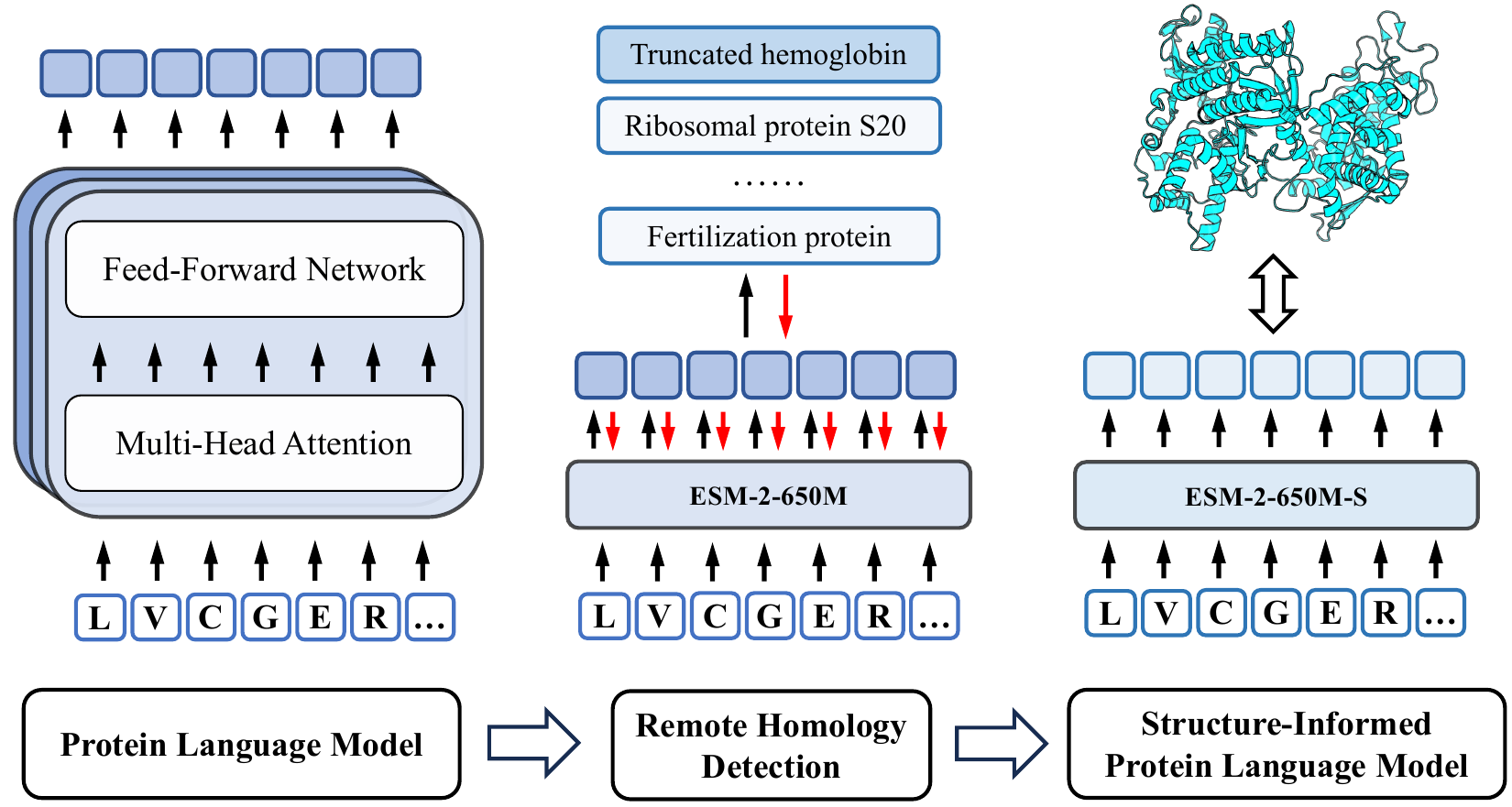}
    \caption{Training Process.}
\end{subfigure}
\hfill
\begin{subfigure}[b]{0.48\textwidth}
    \centering
    \includegraphics[width=\linewidth]{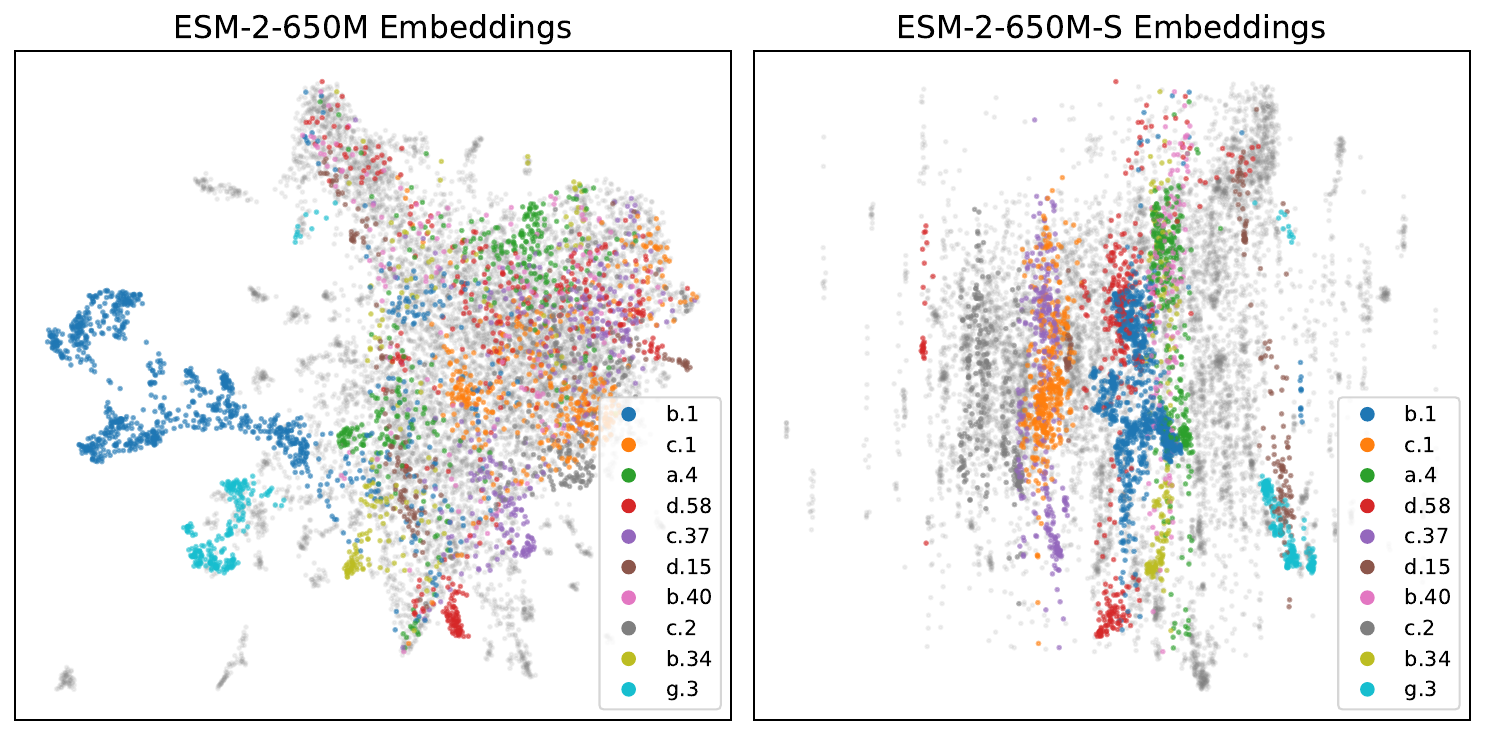}
    \caption{UMAP Embeddings on SCOPe.}
\end{subfigure}
\caption{
Illustration of Training Procedure and Embeddings for Structure-Informed Protein Language Models. (A) Protein language models like ESM-2-650M are enhanced with structural information through training on remote homology detection tasks. This process results in the structure-informed model, ESM-2-650M-S, whose embeddings  represent more structural characteristics. (B) We present UMAP embeddings of both ESM-2-650M and ESM-2-650M-S on the SCOPe dataset. After targeted training, ESM-2-650M-S embeddings show improved separability for different protein folds.}
\label{fig:fold_esm}
\end{figure}


\textbf{Proteins.}
Proteins are made up of amino acids, also known as residues, that form chains via peptide bonds.
There are 20 standard types of residues, and their varied combinations lead to the vast variety of proteins in nature.
The specific arrangement of these residues is crucial in determining the three-dimensional coordinates of every atom in the protein, thus shaping what is known as protein structure.
Protein structures are known to be a direct determinant of protein functions.
In this study, we focus on learning representations based on protein sequences. 
These sequences are denoted as $\gR=[r_1,r_2,\cdots,r_{n}]$, where each $r_i\in\{1,...,20\}$ corresponds to the type of the $i$-th residue.

\textbf{Protein Language Models.}
To effectively encode protein sequences, recent research has treated them as the "language of life", employing methods from large pre-trained language models. This approach aims to capture evolutionary patterns across billions of protein sequences using self-supervised learning. A notable instance of this is the transformer-based protein language model ESM~\citep{rives2021biological,lin2023evolutionary}. This model takes residue type sequences as input, integrating several self-attention layers and feed-forward networks to model dependencies among residues.
These models are pre-trained with a masked language modeling (MLM) loss, which involves predicting the type of a masked residue based on its surrounding context. An additional linear head uses the final-layer representations for this prediction. The loss function for each sequence is defined as:
\begin{align}
    \gL_{MLM} = \begin{matrix}-\E_M [\sum\nolimits_{i\in M} \log p(r_i|r_{/M})]\end{matrix}.
\end{align}
Here, a randomly chosen set of indices $M$ is used for masking, replacing the true token at each index $i$ with a mask token. 
The model's objective is to minimize the negative log likelihood of the correct residue $r_i$, using the masked sequence $r_{/M}$ as context. By leveraging vast amounts of unlabeled data, these models have set new benchmarks in a variety of protein-related tasks~\citep{lin2023evolutionary}.

\textbf{Injecting Structural Information via Protein Remote Homology Detection.}
Protein language models are known to reflect sequence similarity, thereby facilitating the identification of protein homology – the shared ancestry in the evolutionary history of life~\citep{rives2021biological}. 
The challenge in this field, however, goes beyond simple homology detection.
Over the course of natural evolution, protein structures and functions tend to be more conserved than their sequences~\citep{pal2006integrated,liu2014combining}. This means proteins with similar structures and functions might exhibit low sequence identities. Therefore, in the realm of protein homology detection, it is relatively straightforward to identify homologs with high sequence identity, but far more challenging to detect those with low sequence identity.
This specific task of identifying homologous proteins that share structural and functional similarities but differ significantly in sequence is termed \emph{protein remote homology detection}.
The ability to detect remote homologous proteins is crucial in various fields, including proteomics~\citep{kim2014draft} and biomedical sciences~\citep{standley2008protein}.

While~\citet{rives2021biological} has shown that masked language modeling pre-training enables ESM representations to capture remote homology information, these protein language models do not inherently process protein structures as input, nor are they trained with any specific structural loss. To explicitly incorporate structural information into these models, we now take the step to fine-tune the ESM model specifically for the task of protein remote homology detection. 

We employ the remote homology detection dataset from ~\citet{hou2018deepsf}, which is derived from SCOPe 1.75~\citep{murzin1995scop}. 
This dataset is composed of genetically distinct domain sequence subsets that share less than 95\% identity. 
It includes a total of 12,312 proteins, categorized into 1,195 distinct folds, where proteins within the same fold exhibit similar structure patterns. 
To adapt our protein language model for this task, we fine-tune it to take protein sequences as input and attach an MLP head for predicting the fold class label of each protein.
Formally, our objective is to train a protein language model with parameters $\phi$ through training on a protein database $\gR_D$ with corresponding fold labels $c_D$. The optimization involves maximizing the log likelihood:
\begin{equation}
\setlength{\abovedisplayskip}{4pt}
\setlength{\belowdisplayskip}{4pt}
\label{eq:fold}
    \max\nolimits_\phi\; \log p_\phi(c_D|\gR_D) = \begin{matrix}\sum\nolimits_{n\in D} \sum_{c} [c_{n}=c] \log p_\phi(c_{n}=c|\gR_n)\end{matrix}.
\end{equation}
Through this training process, we aim to enhance the model's ability to generate similar representations for proteins that belong to the same fold, thereby injecting structural information.
The high-level idea and effect of structural information injection is shown in Fig.~\ref{fig:fold_esm}.
In practice, limited by the computation resources, we only consider ESM-2-\{8,35,150,650\}M for illustration and exclude ESM-2-3B and ESM-2-15B.
The models are trained on the dataset for 50 epochs using the Adam optimizer with a batch size of 8. To preserve the pre-trained representations, we set the learning rate for ESM to 1e-5, while the learning rate for the prediction head is set to 1e-4.
Models after training on remote homology detection are denoted with a suffix "-S".


\section{Experiment}
\label{sec:exp}

In this section, we aim to evaluate the impact of incorporating structural information by comparing protein language models with structure-informed models across various protein function prediction tasks. Our experiments are run upon ESM-2-\{8,35,150,650\}M, which serves as our protein sequence feature extractor. We explore two approaches to leverage these features: (1) feeding them into a 2-layer Multilayer Perceptron (MLP) predictor (see Sec.~\ref{sec:exp:predictor}) and (2) utilizing them as a similarity metric to retrieve proteins with similar characteristics (see Sec.~\ref{sec:exp:retriever}).
Predictor-based methods aim to determine whether the representations are easily distinguishable for different functions, while retriever-based methods explore whether structural similarity aids in determining protein functions.

\subsection{Evaluation with Predictor-based Methods}
\label{sec:exp:predictor}

\begin{figure*}[t]
    \centering
    \includegraphics[width=\linewidth]{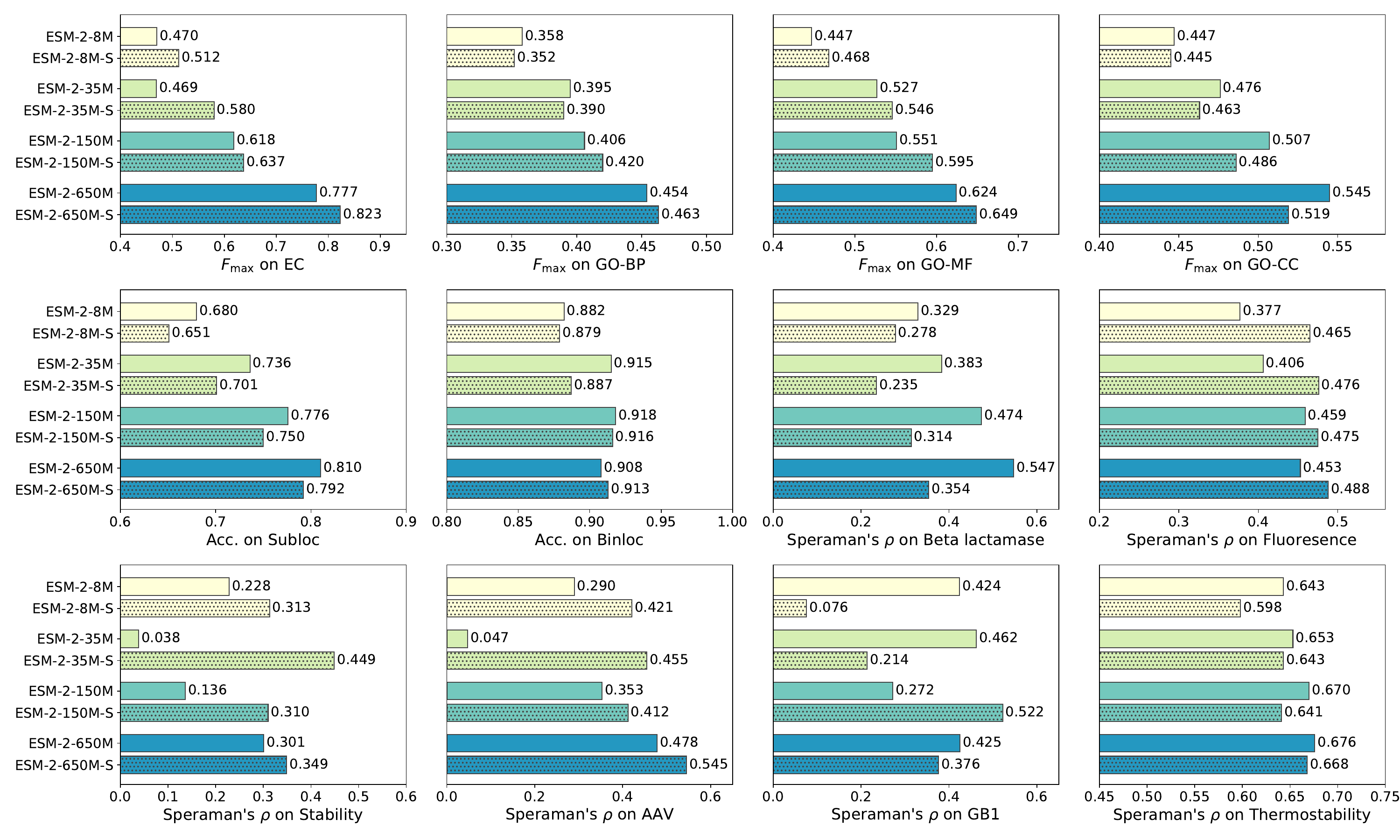}
    \vspace{-2em}
    \caption{Results on function prediction tasks with various sizes of ESM-2 models as feature extractors. Structure-informed models are denoted with suffixes "-S" and highlighted with dots.}
    \label{fig:esm_seq}
    \vspace{-1em}
\end{figure*}

\textbf{Setup.}
We evaluate the methods using three different categories of function prediction tasks.
The first category is function annotation tasks in~\citet{gligorijevic2021structure}, \emph{i.e.}, {Enzyme Commission (EC) prediction} and {Gene Ontology (GO) prediction}. 
EC prediction aims to determine whether a protein can catalyze a biochemical reaction, while GO prediction aims to identify a protein's involvement in specific molecular functions (MF), biological processes (BP), and cellular components (CC).
We split them based on sequence identity cutoff, with the test sets containing sequences that have no more than 95\% similarity to the training set.
The evaluation of performance is based on the protein-centric maximum F-score, denoted as F\textsubscript{max}~\citep{radivojac2013large}.

In addition to function annotation tasks, we also include protein localization prediction tasks, which are related to the in vivo functionality of proteins.
For this evaluation, we utilized two datasets from~\citet{deeploc}: Subcellular localization and Binary localization prediction. These tasks aim to predict the cellular location of natural proteins and are measured by accuracy.

Next, we select a subset of mutation-based tasks from~\citet{xu2022peer} and~\citet{dallago2021flip}. 
These include Beta-lactamase activity~\citep{envision}, Fluorescence~\citep{fluorescence}, Stability~\citep{stability}, AAV fitness~\citep{bryant2021deep}, GB1 fitness~\citep{wu2016adaptation} and Thermostability~\citep{jarzab2020meltome}.
The datasets are split based on the number of mutations and measured by Spearman's correlation.
Please refer to the original papers for more details.

\textbf{Results.}
We freeze the protein language model encoders and feed their outputs into a two-layer MLP head for prediction.
The MLP head are trained for 100 epochs on each task.
The results are reported in Fig.~\ref{fig:esm_seq}.
From the figure, it can be observed that the improvement or decline brought by structurally training are consistent across different sizes of protein language models.
We can gain insights from the results based on different types of tasks.
Firstly, structure-informed ESMs consistently outperform vanilla ESMs in function annotation tasks such as EC, GO-BP, and GO-CC. This can be attributed to the fact that protein structures directly determine their functions, such as catalysis. 
However, for tasks related to cellular location like GO-CC, Subloc, and Binloc, it can be observed that the models perform worse after incorporating structure information. This is likely because protein structures have little influence on where proteins perform their functions. 
As for the remaining tasks based on protein mutants, whether structure-informed models provide benefits depends on whether sequence-based evolutionary information plays a more crucial role than structural information in determining the functions.
In summary, it is important to consider the relationship between protein structures and targeted properties when applying structure-informed training for function prediction.


\subsection{Evaluation with Retriever-based Methods}
\label{sec:exp:retriever}

\begin{figure*}[t]
    \centering
    \includegraphics[width=\linewidth]{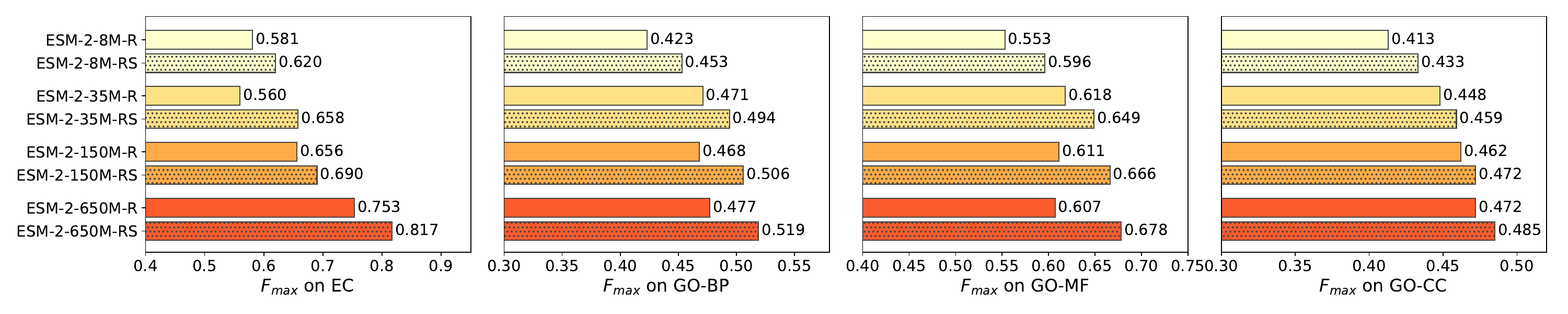}
    \vspace{-2em}
    \caption{F\textsubscript{max} on function annotation with various sizes of ESM-2 models as retrievers with suffixes "-R". Structure-informed retrievers are denoted with suffixes "-RS" and highlighted with dots.}
    \label{fig:esm}
    \vspace{-1em}
\end{figure*}

Besides predictor-based methods, another approach for protein function annotation is to annotate function labels based on labels from similar proteins.
In this subsection, we evaluate the capability of ESM and structure-informed ESM for measure protein similarity for function annotation. 

We first focus on the EC and GO prediction tasks previously discussed. To annotate an unseen protein within the test set, we calculate the cosine similarity between the language model representations of proteins and select the top-5 proteins with the highest similarity scores. Subsequently, we determine the probability of each label by averaging the retrieved proteins' labels, weighted by their similarity scores. The results are presented in Fig.~\ref{fig:esm}.
Notably, when utilizing structure-informed training as retrievers, we observe consistent improvement across all tasks and model sizes. This finding further emphasizes the impact of protein structures in determining their functions.

\begin{wrapfigure}{R}{0.57\textwidth}
\begin{minipage}[b]{0.55\textwidth}
    \vspace{-1.2em}
    \centering
    \includegraphics[width=\linewidth]{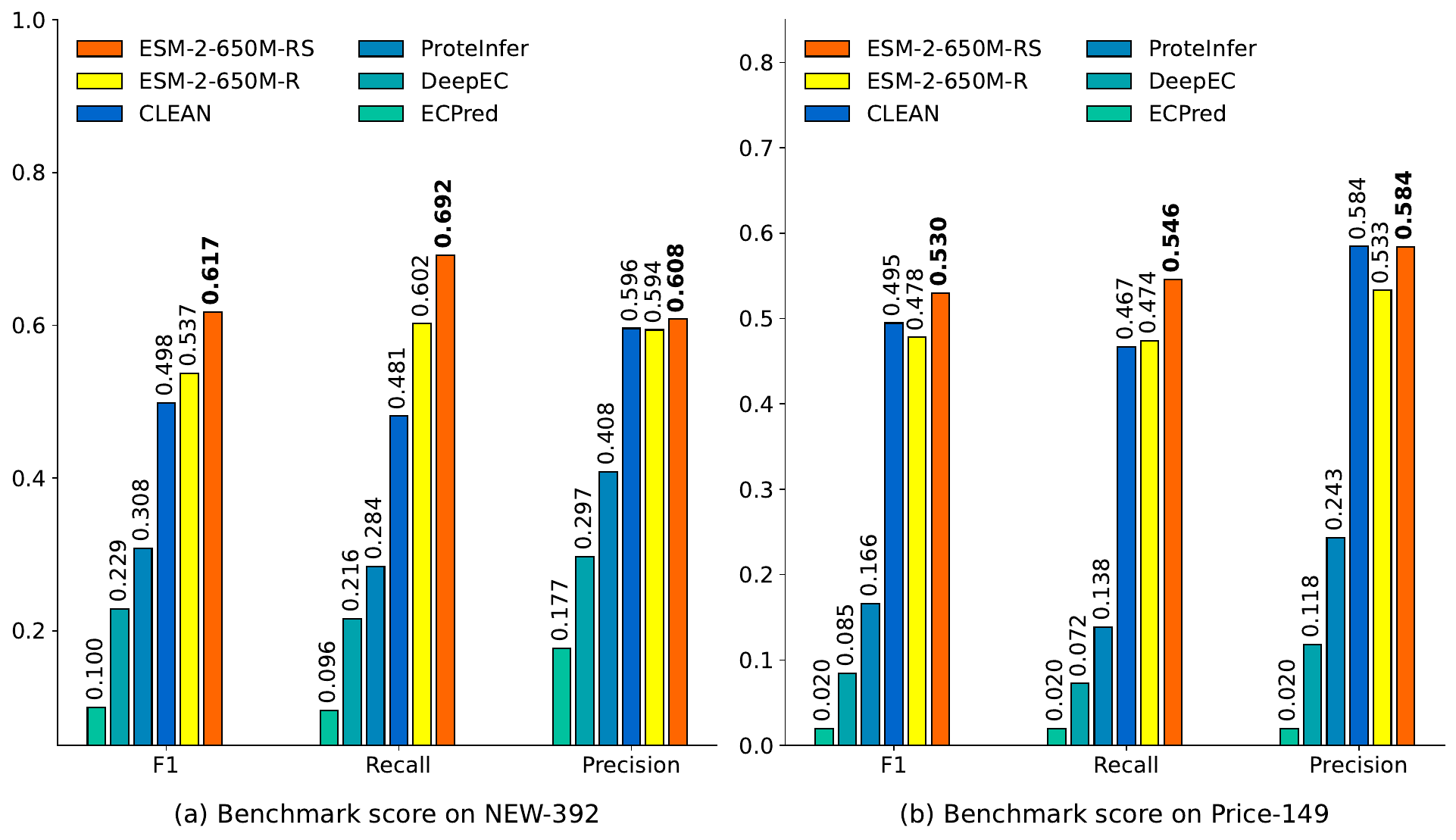}
    \vspace{-1.5em}
    \caption{Results of EC annotation on NEW-392 and Price-149 test sets. Two proposed retrievers are in warm colors, whereas other baselines are in cold colors.
    }
    \label{fig:clean}
    \vspace{-1em}
\end{minipage}
\end{wrapfigure}

In addition, we also explore studies that test EC number annotation under more realistic and challenging settings~\citep{yu2023enzyme,sanderson2021proteinfer}. 
We utilize the Swiss-Prot dataset collected in~\citet{yu2023enzyme} with 227,363 protein sequences, as the retrieval dataset.
We then test various retriever-based methods on two independent test sets.
The first, an enzyme sequence dataset, includes 392 sequences that span 177 different EC numbers. These sequences were released after April 2022, which reflects a real-world scenario where the functions of the query sequences are still unknown.
The second test set, known as Price-149, is a benchmark dataset curated by~\citet{sanderson2021proteinfer}. It consists of experimentally validated findings from the study by~\citet{price2018mutant}. This dataset includes sequences that were previously mislabeled or inconsistently annotated in automated systems, making it a challenging benchmark for evaluation.
To establish baselines, we consider four EC number prediction tools: CLEAN~\citep{yu2023enzyme}, ProteInfer~\citep{sanderson2021proteinfer}, ECPred~\citep{dalkiran2018ecpred} and DeepEC~\citep{ryu2019deep}. The results of these tools are directly taken from the CLEAN paper by~\citep{yu2023enzyme}.
For comparison, we test the performance of two neural retrievers introduced in our paper : ESM-2-650M-R and ESM-2-650M-RS.

The results are plotted in Fig.~\ref{fig:clean}. 
Both our retrievers surpass the performance of CLEAN on the NEW-392 test set in F1 score, despite not undergoing any supervised training on the training set, a process that CLEAN underwent. This underscores the potency of protein language models.
Furthermore, the strategy of integrating structural insights into ESM proves to be effective for EC number prediction, with observable enhancements on both test sets. Specifically, on the more challenging Price-149 set, while CLEAN slightly outperforms ESM-2-650M, it falls short against ESM-2-650M when structural information is incorporated. 
This reaffirms the significance of structural similarity in function similarity assessments.
To conclude, structure-informed training continues to demonstrate potential in practical function annotation scenarios, emphasizing the critical role of modeling structural similarities between proteins.

\section{Conclusion}

In this study, we investigate the integration of remote homology detection tasks for infusing structural information into protein language models. Our experimental findings indicate that incorporating structural information leads to consistent enhancement in function annotation accuracy. Nonetheless, it remains crucial to consider the connection between protein structures and targeted properties when applying structure-informed training. We envision that this study paves the way for further exploration of structural distillation techniques to enhance protein language models.

\section*{Acknowledgments}


This project is supported by AIHN IBM-MILA partnership program, the Natural Sciences and Engineering Research Council (NSERC) Discovery Grant, the Canada CIFAR AI Chair Program, collaboration grants between Microsoft Research and Mila, Samsung Electronics Co., Ltd., Amazon Faculty Research Award, Tencent AI Lab Rhino-Bird Gift Fund, a NRC Collaborative R\&D Project (AI4D-CORE-06) as well as the IVADO Fundamental Research Project grant PRF-2019-3583139727.

\bibliography{iclr2024_conference}
\bibliographystyle{iclr2024_conference}


\end{document}